# Why the overlap formula does not lead to chiral fermions[*]

Maarten Golterman[a] and Yigal Shamir[b]

[a]Department of Physics, Washington University,
St. Louis, MO 63130-4899, USA

[b]School of Physics and Astronomy, Beverly and Raymond Sackler Faculty of Exact Sciences,
Tel-Aviv University, Ramat Aviv 69978, Israel

We describe a conceptually simple, but important test for the overlap approach to the construction of lattice chiral gauge theories. We explain the equivalence of the overlap formula with a certain waveguide model for a simple set of gauge configurations (the trivial orbit). This equivalence is helpful in carrying out the test, and casts serious doubts on the viability of the overlap approach. A recent note by Narayanan and Neuberger which points out a mistake in our previous work is irrelevant in this context.

## 1. Introduction

In almost all attempts to construct chiral gauge theories on the lattice thus far gauge invariance is explicitly broken on the lattice. The hope has always been that a gauge invariant chiral gauge theory would be recovered in the continuum limit for fermion representations that are free of anomalies in the gauged symmetries [1]. An important consequence of this gauge noninvariance is the fact that, in such proposals, the fermions couple to the gauge degrees of freedom (the degrees of freedom along the gauge orbits), as well as to the physical components of the gauge fields. This leads in many cases to unexpected nonperturbative dynamics, and may change the fermion spectrum of the model.

A simple example is the usage of the Wilson mass terms for fermions chirally coupled to gauge fields, pioneered in ref. [2]. If the lefthanded fermion field $\psi^-(x)$ transforms to $V(x)\psi^-(x)$ under a gauge transformation ($V$ is an element of the gauge group) while the righthanded field $\psi^+(x)$ is neutral, the Wilson term transforms as

$$\frac{r}{2}\sum_\mu \left(\overline{\psi}(x)\psi(x+\mu) - \overline{\psi}(x)\psi(x) + \text{h.c.}\right) \to$$
$$\frac{r}{2}\sum_\mu \left(\overline{\psi}^+(x)V(x+\mu)\psi^-(x+\mu) + \ldots\right). \quad (1)$$

[*]presented by MG

Clearly, the Wilson parameter $r$ turns into a Yukawa-like coupling, and the fermions couple to the gauge degrees of freedom $V(x)$.

This observation leads to an important test for any of the proposals in which gauge invariance is explicitly broken on the lattice. If we choose the lattice gauge fields to be pure gauge, $U_\mu(x) = V(x)V^\dagger(x+\mu)$, and integrate over $V$, we should obtain free, undoubled fermions in the continuum limit, with the correct coupling to the transverse gauge fields when they are turned back on. This test has failed for all those attempts for which it has been carried out [3], in particular [4] for the waveguide implementation of Kaplan's domain wall fermions [5]. It is important to investigate the same issue for the overlap approach [6].

Here we report on work [7] in which we showed that for this restricted set of gauge fields, the overlap approach is identical to a modified waveguide model. This has consequences for the fermion spectrum similar to what happens in the original waveguide approach, and casts very strong doubts on the viability of the overlap approach for defining lattice chiral gauge theories. Our paper does contain a technical mistake for the case of topologically nontrivial gauge fields (an erratum will appear) [8], but since our simple test does not involve such gauge fields, this observation does not invalidate our conclusions about the fermion spectrum. We maintain that the overlap approach is in serious trouble.

## 2. Waveguide approach

Here we do not have the space to give full details of the definition of the various models, see ref. [7]. The original waveguide model has a fermion action

$$\begin{aligned}S_F &= \sum_{s\in\text{WG}} \overline{\psi}_s \left(\slashed{D}(U) - W(U) + m(s)\right)\psi_s \\ &+ \sum_{s\notin\text{WG}} \left(\overline{\psi}_s(\slashed{\partial} - W + m(s))\psi_s - \sum_s \overline{\psi}_s\psi_s\right) \\ &- \sum_{s\neq L,-L} \left(\overline{\psi}_s^- \psi_{s+1}^+ + \overline{\psi}_{s+1}^+ \psi_s^-\right) \\ &- y\left(\overline{\psi}_L V^\dagger \psi_{L+1}^+ + \text{h.c.}\right) \\ &+ \text{similar at } s=-L. \end{aligned} \quad (2)$$

Here $W$ ($W(U)$) denotes the (gauged) 4-d Wilson term. $s$ is the coordinate in the fifth direction, and we take $-2L+1 \leq s \leq 2L$. $WG$ is the waveguide, namely, the range $-L+1 \leq s \leq L$ centered around the domain wall where the mass changes sign. We choose antiperiodic boundary conditions in the 5th direction, and there is an antidomain wall outside the waveguide. The gauge field $U$ is purely four dimensional. $y$ is a Yukawa coupling which we will choose to be one for most of what follows. This model can be obtained by first choosing $V = 1$ (in which case it is not gauge invariant), and then performing a gauge transformation $U_\mu(x) \to V(x)U_\mu(x)V^\dagger(x+\mu)$, $\psi_s \to V\psi_s$ inside the waveguide (the fermions outside the waveguide are not coupled to the gauge field and do not transform), leading to the action above. The fermions clearly couple to the gauge degrees of freedom, $V$. We can now perform our test by setting $U = 1$ and integrating over $V$.

The outcome [4] can be summarized as follows. For smaller or larger values of $y$ two distinct symmetric phases exist ($\langle V \rangle = 0$). There is a righthanded massless fermion bound to the domain wall (inside the waveguide), and a lefthanded one bound to the antidomain wall (outside the waveguide). New defects are introduced at the waveguide boundaries *dynamically* by the vanishing of $\langle V \rangle$, and a lefthanded (righthanded) massless fermion appears at the inside (outside) of one of the waveguide boundaries. (For the large $y$ symmetric phase many more mirror fermions appear at the waveguide boundaries.) Both the domain wall fermion and the one inside the waveguide boundary will couple to the gauge field, rendering the theory vectorlike. For intermediate values of $y$ the symmetry is broken and the massless fermions at the boundary combine into a massive Dirac fermion, with a mass set by $\langle V \rangle$, like the gauge field mass. The situation is essentially the same as in the mirror fermion model [9]. It is important to note that the dynamics of the gauge degrees of freedom ($V$) plays an essential role in these conclusions. It teaches us that without considering this dynamics, one cannot decide on the success of any gauge noninvariant proposal for the construction of a lattice chiral gauge theory. This clearly includes the overlap approach.

## 3. Transfermatrix

Before we get to the overlap formula, we first introduce a transfermatrix representation of the partition function for the waveguide model. For simplicity we take $y = 1$ and choose the unitary gauge, $V = 1$. We take the time direction to be the fifth direction (labeled by $s$), and construct the transfermatrix following Lüscher [10]. The fermionic partition function takes the form

$$\begin{aligned}Z_F(U) &= \text{prefactor}(U, L) \\ &\times \text{tr}\left(T_-^L T_-(U)^L T_+(U)^L T_+^L\right),\end{aligned} \quad (3)$$

where $T_\pm(U)$ is the transfermatrix inside the waveguide (where the gauge field $U$ is nontrivial) in the region with positive/negative mass, and $T_\pm$ is the same outside the waveguide (where $U = 1$).

Representing $T$ in terms of its eigenvalues $\lambda_n$ and eigenstates $|n\rangle$ as $T = \sum_n |n\rangle\lambda_n\langle n|$, we obtain for large $L$

$$\begin{aligned}Z_F(U) &\sim \text{prefactor}(U, L)\left(\lambda_-\lambda_-(U)\lambda_+(U)\lambda_+\right)^L \\ &\times \langle I-|U-\rangle\langle U-|U+\rangle\langle U+|I+\rangle\langle I+|I-\rangle\end{aligned} \quad (4)$$

(as long as the overlaps appearing in this equation are nonvanishing). In this equation $|U\pm\rangle$ are the groundstates of $T_\pm(U)$, and $|I\pm\rangle$ those of $T_\pm$. $\lambda_\pm(U)$ and $\lambda_\pm$ are the corresponding eigenvalues.

The remaining $L$ dependence can be compensated by the introduction of heavy, gauge invariant Pauli-Villars (PV) fields (no coupling to $V$) as described in detail in ref. [7], and the final result becomes

$$Z = \langle I-|U-\rangle\langle U-|U+\rangle\langle U+|I+\rangle\langle I+|I-\rangle, \quad (5)$$

which is actually valid for all $U$. $Z$ is zero if the overlaps $\langle U-|U+\rangle$ or $\langle U+|I+\rangle$ vanish ($\langle I-|U-\rangle$ never vanishes [6]). The factors on the righthand side correspond to the first waveguide boundary, the domain wall, the second waveguide boundary and the antidomain wall respectively.

## 4. The overlap

The overlap formula for the partition function for topologically trivial configurations reads [6]

$$Z_o(U) \equiv \frac{\langle I-|U-\rangle\langle U-|U+\rangle\langle U+|I+\rangle}{|\langle I-|U-\rangle||\langle U+|I+\rangle|}. \quad (6)$$

The numerator of this expression is clearly identical to eq. (5), up to a trivial ($U$-independent) factor, and the question arises whether we can also get the denominator from a euclidean path integral. The answer is that this can indeed be done for a theory with an even number of same chirality flavors. Here we will discuss the case of $n = 4$ flavors, for which the overlap formula reads

$$\frac{(\langle I-|U-\rangle\langle U-|U+\rangle\langle U+|I+\rangle)^4}{(\langle I-|U-\rangle\langle U-|I-\rangle\langle I+|U+\rangle\langle U+|I+\rangle)^2}. \quad (7)$$

The idea is to use different, gauge noninvariant PV fields [7]. We introduce two complex bosonic spinor fields $\tilde\phi_\pm$ in the same representation of the gauge group as the fermions. Again they couple to the gauge fields only inside the waveguide, and therefore also to $V$ at the waveguide boundaries. We take the PV action for each to be quadratic:

$$S_{PV} = \sum_{s,s',\pm} \tilde\phi^\dagger_{\pm s}(D^\dagger D)_{s,s'}\tilde\phi_{\pm s'}, \quad (8)$$

where $D$ is the fermion Dirac operator where however we take $m(s) = +m$ everywhere for $\tilde\phi_+$, and $m(s) = -m$ for $\tilde\phi_-$ (for details see ref. [7]). This implies that there are no (anti)domain walls for the PV fields, but because they couple in essentially the same way to $V$ as the fermions, they will feel the dynamically generated defects. This is also exhibited by the expression of the partition function for these fields:

$$Z_{PV\pm} = \left(\text{prefactor} \times \text{tr}\left(T_\pm(U)^{2L}T_\pm^{2L}\right)\right)^{-2} \to \quad (9)$$
$$\left(\text{prefactor} \times (\lambda_\pm(U)\lambda_\pm)^{2L}\langle I\pm|U\pm\rangle\langle U\pm|I\pm\rangle\right)^{-2}.$$

The minus sign in the exponent comes from the bosonic statistics, and the 2 from the $D^\dagger D$ structure of the kinetic operator for the PV fields. Combining eqs. (4,9) (and including a normalization factor $\langle I+|I-\rangle^{-4}$) leads exactly to the overlap formula, eq. (7). This result is valid in the case that the numerator and denominator of eq. (7) do not vanish. In particular, it is true for $U$ pure gauge where, moreover, convergence in the limit $L \to \infty$ is uniform. (This result is not valid for topologically nontrivial gauge fields [6,8], erratum to [7], but that is irrelevant here.)

To summarize, for the gauge configurations that we are interested in, we have established an exact equivalence between the overlap formula and the $L \to \infty$ limit of a modified waveguide model, where the modification of the waveguide model consists of the gauge noninvariant PV sector. The above analysis can easily be extended to values of the Yukawa coupling $y$ other than one, following techniques used in ref. [11].

## 5. Spectrum

The techniques used in order to investigate the spectrum of the original waveguide model for the restricted set of gauge fields $U_\mu(x) = V(x)V^\dagger(x+\mu)$ are also useful in determining the spectrum of the modified waveguide model (for both the fermion and the PV fields), and therefore the overlap for this set of gauge fields. We will describe here briefly what one would expect.

Let us assume we are in a symmetric phase, which is where one wants to be in order to construct an asymptotically free chiral gauge theory. This can certainly be arranged at small enough $y$, and there one expects the massless fermion spectrum to be that of the small $y$ region described in sect. 2, except that it is now quadru-pled. There will be no massless PV modes at the (anti)domain wall, but there will be at both waveguide boundaries for $\tilde{\phi}_+$: on one side there will be two lefthanded and on the other side there will be two righthanded massless modes just inside the waveguide, and similar modes just outside the waveguide. (There are no zeromodes in $\tilde{\phi}_-$.) Again, if we turn on the full gauge field, all massless modes inside the waveguide will couple to the gauge field, whereas those outside will not.

Such a spectrum exhibits two disasters: one is that this spectrum is entirely vectorlike, the second being the presence of massless modes of the wrong statistics. Indeed, there is no guarantee that the overlap approach satisfies unitarity.

One might want to employ the overlap approach for describing QCD with for instance four flavors. One would get the partition function for this theory by multiplying eq. (7) by its complex conjugate, obtaining

$$(\langle U-|U+\rangle\langle U+|U-\rangle)^4, \qquad (10)$$

and one sees that all overlaps corresponding to the waveguide boundaries disappear from this expression. This is indeed consistent with the spectrum described above: taking the complex conjugate corresponds to interchanging all right- and lefthanded modes and adding that to the unchanged spectrum. In this case it is easy to convince oneself that the effect of all massless mirror fermions (*i.e.* those living at the waveguide boundaries) and all massless PV modes cancel between each other.

## 6. Conclusion

We showed how an important test can be applied to the overlap approach to the construction of lattice chiral gauge theories by establishing equivalence with a modified waveguide model for a sufficiently large class of gauge configurations.

The test itself concerns the fermion spectrum of the theory directly, and not only the feedback of fermion (or PV) fields on the gauge sector. While the latter is interesting and important in itself, knowing the complete fermion spectrum (with the transverse gauge fields turned off, as explained in the introduction) is at least equally important.

The equivalence to a waveguide model makes the overlap approach accessible to techniques developed in the past in order to perform this test. We have described what spectrum we expect based on previous experience with waveguide models. Our conclusion is that there are very serious doubts concerning the viability of the overlap approach.

Lastly, we would like to add that we believe a waveguide model can be constructed which represents the overlap for all gauge fields, including topologically nontrivial ones. We hope to return to this question in the future.

## Acknowledgements

MG is supported in part by the Department of Energy under contract number #DOE-2FG02-91ER40628 and by a DOE Oustanding Junior Investigator grant. YS is supported in part by the US-Israel Binational Science Foundation, and the Israel Academy of Science.